\documentclass[secnumarabic,amssymb, nobibnotes, aps, prl,longbibliography,superscriptaddress, preprint]{revtex4-2}

\usepackage{graphicx}% Include figure files
\usepackage{dcolumn}% Align table columns on decimal point
\usepackage{bm}% bold math
\usepackage{natbib}
\usepackage{bibentry}
\usepackage{xcolor}
\usepackage{amsmath}
\usepackage{mathrsfs}

\begin{document}

\title{Photo-induced Hidden Phase of 1T-TaS2 with Tunable Lifetime }% Force line breaks with \\

\author{Pierre-Adrien Mante}
\email{pierre-adrien.mante@chemphys.lu.se}
\affiliation{Division of Chemical Physics and NanoLund, Lund University, Sweden}

\author{Chin Shen Ong}
\affiliation{Department of Physics and Astronomy, Uppsala University, Sweden}

\author{Daniel Finkelstein Shapiro}
\affiliation{Instituto de Química, Universidad Nacional Autónoma de México, Mexico}

\author{Arkady Yartsev}
\affiliation{Division of Chemical Physics and NanoLund, Lund University, Sweden}

\author{Oscar Gr\aa n\"as}
\affiliation{Department of Physics and Astronomy, Uppsala University, Sweden}

\author{Olle Eriksson}
\affiliation{Department of Physics and Astronomy, Uppsala University, Sweden}
\affiliation{School of Science and Technology, Örebro University, Sweden}

\date{\today}% It is always \today, today,

\begin{abstract}
Phase transitions are ubiquitous, appearing at every length scale from atoms~\cite{sachdev_2011} to galaxies~\cite{10.1016/0370-1573(80)90091-5}. In condensed matter, ultrafast laser pulses drive materials to highly non-equilibrium conditions allowing transitions to new phases of matter not attainable under thermal excitation. Despite the intense scrutiny these hidden phases have received, the details of the dynamics of transition and reestablishment of the ground state remains largely unexplored. Here, we show the transition to a hidden phase of 1T-TaS$_2$ driven by the screening of Coulombic repulsive interaction by photo-excited electrons. The temporal evolution of the coherent lattice dynamics highlights the existence of a novel phase with a laser fluence dependent lifetime. The modeling of the dynamics reveals that the transition is caused by photo-excited carriers and it disappears at the rate of electron-phonon scattering. Our results demonstrate how femtosecond laser absorption leads to a decoupling of the electronic and lattice sub-systems, opening the way to novel states of matter, which can be controlled with light. We expect our investigation to be a starting point towards the development of novel ultrafast photonics devices, such as switches and modulators, taking advantage of fast and tunable phase transitions.
\end{abstract}

\maketitle

Phase change materials have received intense scrutiny in the past decades due to their technological applications from memory devices~\cite{10.1021/cr900040x} to neuromorphic computing~\cite{10.1038/s41578-018-0076-x}. Traditionally, phases of matter are considered in the context of thermal equilibrium, where temperature or pressure are steady-state parameters~\cite{papon2006physics}. However, novel phases of matter may arise due to non-equilibrium between the various sub-systems (electron, lattice, spin...) of a material. In particular, the use of ultrafast lasers as a driving source allows reaching these non-ergodic conditions and driving matter to novel phases~\cite{10.1073/pnas.1808414115,10.1038/nmat2929,PhysRevLett.91.136402,PhysRevLett.107.177402,doi:10.1126/science.1241591,10.1038/nature09539,10.1103/physrevlett.97.067402,10.1126/sciadv.1400173,10.1126/sciadv.1500606,10.1063/5.0052311}. This approach is particularly useful to investigate strongly correlated materials, where the competition between the different phenomena ruling the phase of a material can be modified, leading to the emergence of new phases. Based on this principle, hidden phases of matter, \textit{i.e.} phases of matter that do not appear under equilibrium conditions, have been recently evidenced~\cite{10.1038/nmat2929,doi:10.1126/science.1241591,10.1126/sciadv.1400173}. Nevertheless, due to the complex interplay between the various degrees of freedom during these transitions,identifying the mechanisms at play remains a challenging task.

Layered transition metal dichalcogenides possess a complex phase diagram showcasing superconductivity, Mott insulating states and charge density waves (CDW). As such, they are a perfect playground to investigate the existence of hidden states, as well as their formation and dissipation dynamics. 1T-TaS$_2$ has especially been investigated following the discovery of a laser-induced stable hidden state, with drastically different optical and electrical properties~\cite{doi:10.1126/science.1241591}. This hidden state can be reached at temperature lower than 70~K, where the material usually exhibits a commensurate CDW. In this seminal work, Stojchevska \textit{et al.} highlight the existence of this hidden phase by tracking the coherent amplitude modes of the CDW, which is a fingerprint of the state of matter. Recently, another hidden phase of 1T-TaS$_2$ was experimentally observed~\cite{10.1126/sciadv.1400173}, stemming from the room-temperature phase, where the material a nearly commensurate CDW. However, this phase was not confirmed by other experimental methods, such as the change of the coherent amplitude modes.

Here, we investigate the ultrafast vibrational response of 1T-TaS$_2$ to the absorption of a femtosecond laser pulse at room temperature. We reveal the existence of a coherent amplitude modes distinct from the known spectra of the various phases of this material, and we associate it to a new hidden phase of the material. We then study and model the dynamics of this phase, highlighting its origin: the screening of Coulombic potential by the photoexcited electrons with high kinetic energy. We also show that, as the electrons lose kinetic energy through scattering with phonons, the new phase slowly disappears and the system reverts back to the thermodynamically stable phase. Finally, our model shows that by tuning the driving laser fluence, it is possible to control the lifetime of the hidden phase, thus opening the path to the development of novel ultrafast photonic devices. 

\section{Results}
\begin{figure*}
    \centering
    \includegraphics[width=\textwidth]{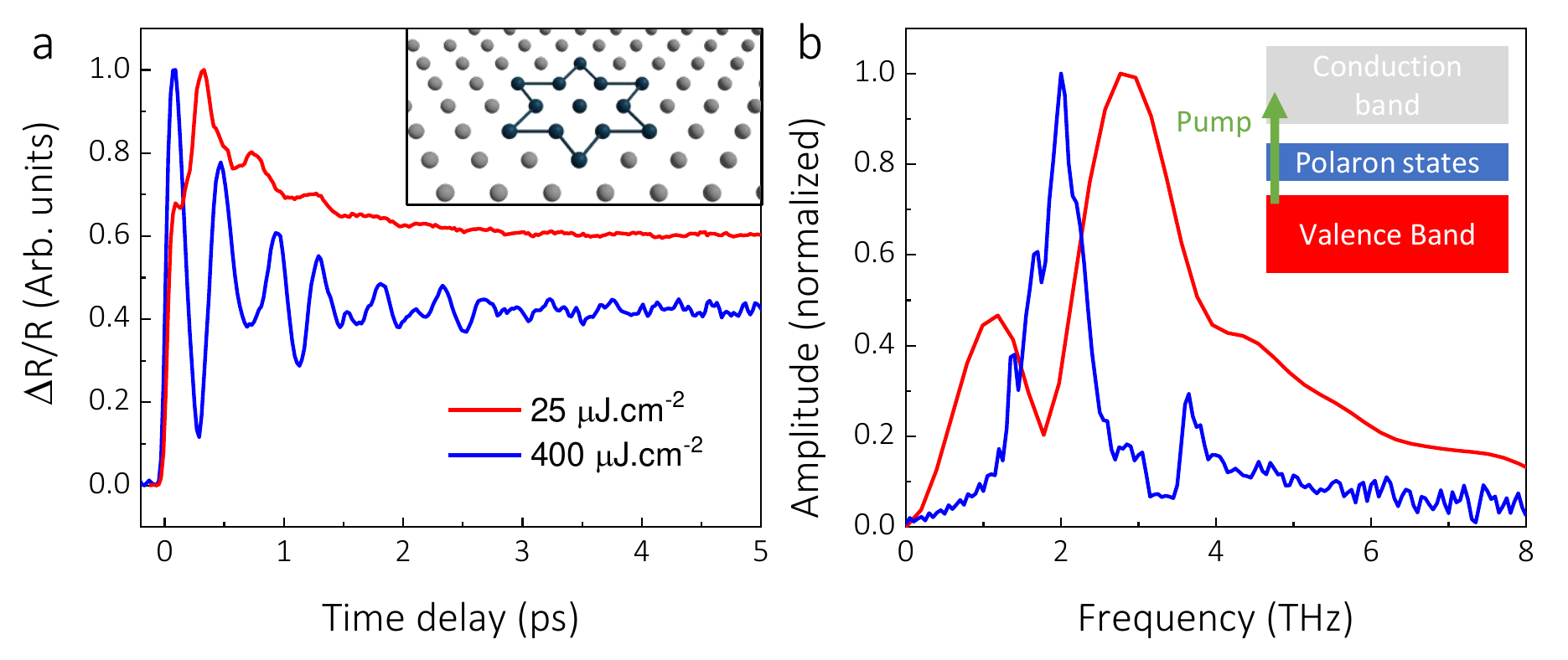}
    \caption{\textbf{The nearly-commensurate and hidden phase of 1T-TaS$_2$} (a) Transient reflectivity obtained on 1T-TaS$_2$ at a pump fluence of 25~$\mu$J.cm$^{-2}$. Inset: 1T-TaS$_2$ lattice showing the formation of a star of David polarons. (b) Fourier transforms of the oscillatory part of the transient reflectivity obtained at 25 and 400~$\mu$J.cm$^{-2}$}
    \label{fig:NCCDW}
\end{figure*}

\textbf{Nearly-commensurate charge density waves.} 1T-TaS$_2$ possesses a rich phase diagram containing numerous charge density waves with different degrees of commensurability between electrons and lattice. Above 550~K, the material is found in the normal phase that displays metallic behavior. Upon cooling, it undergoes a transition to an incommensurate (IC) charge density wave (CDW) until 350~K, where it further transforms into a nearly-commensurate (NC) phase. The formation of the NCCDW is accompanied by a distortion of the lattice structure resulting in the formation of star of David polarons as shown in the inset of Fig~\ref{fig:NCCDW}a. The NC phase is composed of hexagonal domains of star of David polarons surrounded by domain walls. Finally, at a temperature of 180~K, 1T-TaS$_2$ transforms into a fully commensurate (C) phase, forming a superlattice of polarons. The physics of these different CDW is ruled by the Hubbard Hamiltonian:

\begin{equation}
	H=-t \sum_{\langle i, j \rangle, \sigma}(c_{i,\sigma}^{\dagger}c_{j,\sigma}+c_{j,\sigma}^{\dagger}c_{i,\sigma})+U\sum_i n_{i,\uparrow}n_{i,\downarrow}
\end{equation}

where $c_{i,\sigma}^{\dagger}$ and $c_{i,\sigma}$ are the creation and annihilation operators for an electron at location $i$ with spin $\sigma$ and $n_{i,\sigma}=c_{i,\sigma}^{\dagger}c_{i,\sigma}$ is the occupation operator. In this Hamiltonian, a competition exist between kinetic energy, through the hopping integral $t$, and Coulomb repulsion, with the on-site interaction strength $U$. Electrons at the center of the star of David are thus localized due to the dominance of the on-site interaction. 

The lattice distortion associated with the star of David polarons leads to the appearance of amplitude and zone-folded acoustic modes. These vibrational modes are direct signatures of the phase of the material and can be investigated using ultrafast pump-probe spectroscopy~\cite{doi:10.1126/science.1241591}. Figure~\ref{fig:NCCDW}a shows the transient reflectivity obtained for a pump wavelength of 550~nm and a probe wavelength of 700~nm. The inset of Fig.~\ref{fig:NCCDW}b highlights that for a 550~nm pump, valence band electrons, from mostly S orbitals~\cite{10.1103/physrevb.98.195425}, are brought to the conduction band where they can screen the on-site interaction strength. In this experiment, the pump fluence was limited to 25~$\mu$J.cm$^{-2}$, corresponding to a temperature increase of 3~K, much lower than the NC-to-IC transition temperature. This low fluence allows to minimize the perturbation of the system, and thus extract the intrinsic properties of the NC phase. After an initial rise in the reflectivity due to the photoexcitation of carriers, the signal decays due to the thermalization of the electronic system with the lattice on a time scale of 300~fs~\cite{10.1038/nature09539,10.1103/physrevlett.97.067402}. Superimposed on this decay are strong oscillations caused by the excitation of coherent phonons and amplitude modes.

The Fourier transform of the oscillatory part of the signal, diplayed in Fig.~\ref{fig:NCCDW}b, shows a series of peaks located at 1.7, 1.9, 2.1, 2.4, and 3.6~THz. These frequencies are in good agreement with experiments performed with Raman~\cite{10.1016/0378-4363(81)90284-9} and ultrafast spectroscopy~\cite{PhysRevB.94.115122}. The mode at 2.1~THz is the amplitude mode of the CDW, while the modes at 1.7, 1.9 and 2.4~THz have been attributed to zone-folded acoustic phonons. The origin of the mode at 3.6~THz has been scarcely discussed in the case of 1T-TaS$_2$~\cite{PhysRevB.94.115122}, but by drawing a parallel with other layered transition metal dichalcogenide, such as 1T-TaSe$_2$ and 1T-TiSe$_2$~\cite{PhysRevLett.91.136402}, and due to the softening of this mode at the phase transition between the C and NC phase, we attribute this frequency to another amplitude mode of the CDW.

\textbf{Photoinduced phase transition.} Figure~\ref{fig:NCCDW}a and b also shows the transient reflectivity and Fourier transform of the signal obtained for a pump fluence of 400~$\mu$J.cm$^{-2}$, respectively. The frequencies of the modes are shifted to higher frequencies, namely 3 and 4.5~THz, indicative of a modification of the lattice, and potentially of the phase of the material. However, there may be multiple origins to this change of the coherent dynamics. In the following, we rule out some of these possibilities.

\begin{figure*}[!h]
    \centering
    \includegraphics[width=\textwidth]{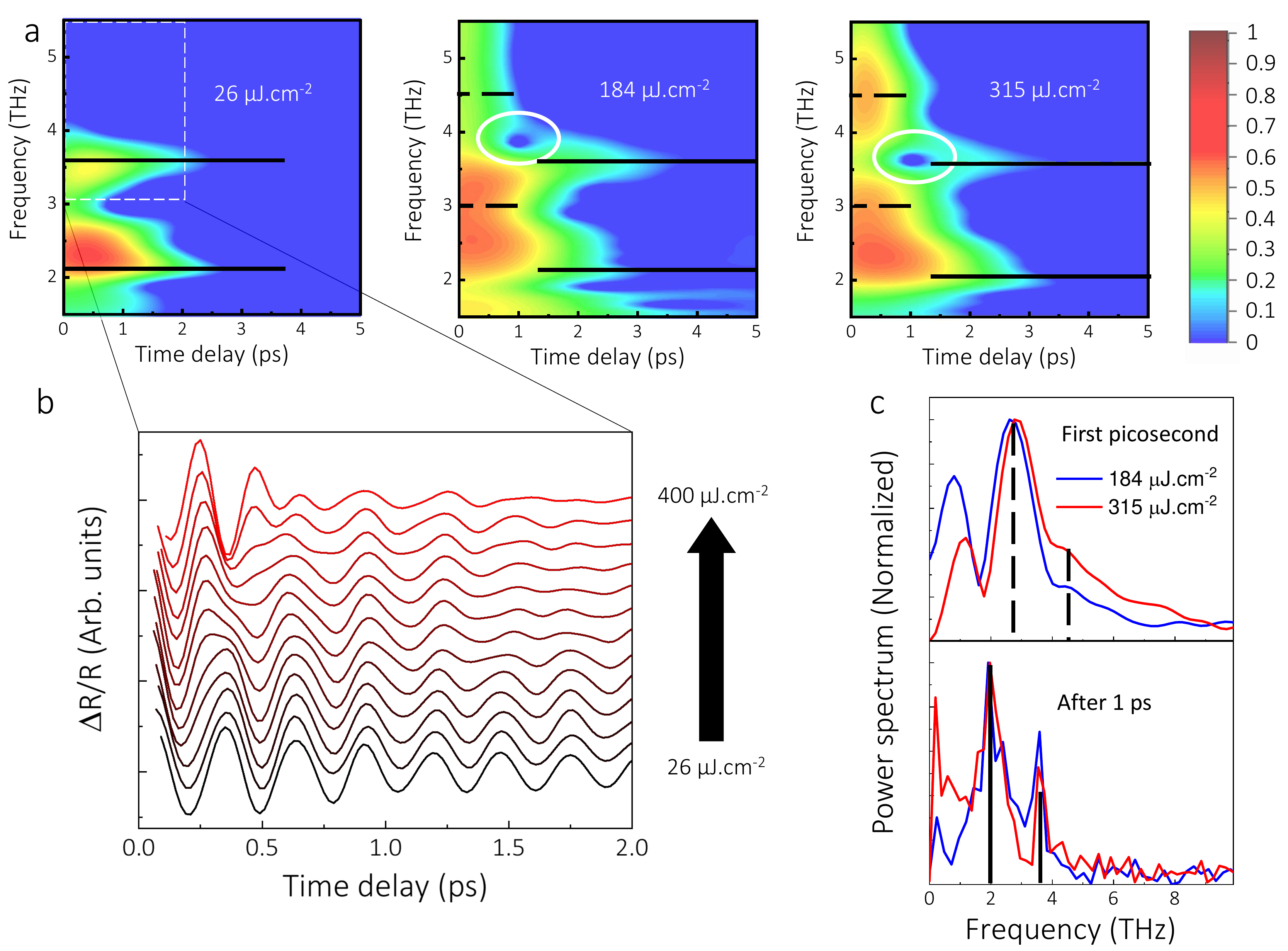}
    \caption{\textbf{Fluence dependence of the phase transition.} (\textbf{a}) Time frequency analysis of the oscillatory components of the transient reflectivity obtained for a pump fluence of 25, 190 and 330~$\mu$J.cm$^{-2}$. The solid black lines shows the main modes' frequencies of the NCCDW phase, the dashed black lines represents the frequencies of the modes of the new phase, and the white circle indicates the destructive interferences occuring at the transition between the two phases. (\textbf{b}) Oscillatory components of the transient reflectivity for pump fluence ranging from 25 to 330~$\mu$J.cm$^{-2}$ filtered from 3 to 6~THz. \textbf(c) Fourier transform of the first picosecond of the oscillatory components of the transient reflectivity obtained at 190 and 330~$\mu$J.cm$^{-2}$ showing the coherent vibrations associated with the new phase (upper panel, dashed lines). Fourier transform of the same signals but for time delays larger than 1~ps highlighting the frequencies of the NCCDW (lower panel, solid lines).}
    \label{fig:TFA}
\end{figure*}

The thermally induced NC to IC phase transition could be responsible for the observed changes in frequencies, however multiple considerations allow to rule out this possibility. First, the temperature increase induced by the pump is calculated to be at the most 40~K, which is not sufficient to reach 350~K, the temperature of the NC to IC phase transition (see supplementary note 1 for detailed calculation of the lattice temperature increase). Second, the oscillation appears on an ultrafast time scale, on the order of 100~fs, which indicates that the phase transition is of electronic origin and not due to a temperature increase of the lattice and a subsequent structural transition. Finally, we performed experiments at a temperature of 370~K, above the NC to IC transition temperature, and we observed a single frequency at 2.5~THz, lower than what we observe in the new phase, in agreement with Raman studies~\cite{10.1016/0378-4363(81)90284-9}. We conclude that the frequencies we are observing do not corresponds to the eigenmodes of the IC phase. 

Another possibility is the anharmonicity of the vibration in 1T-TaS$_2$~\cite{10.1103/physrevb.83.035104}, which could explain the shift we observe at high power. To discriminate between a new phase or anharmonic behavior,  we investigate the evolution of the vibrational content of the transient reflectivity for an increasing pump power. We perform experiments with a pump fluence ranging from 25 to 400~$\mu$J.cm$^{-2}$ for a pump wavelength of 550~nm and a probe wavelength of 700~nm. Figure~\ref{fig:TFA}a shows the time-frequency analysis of the oscillatory components of the transient reflectivity obtained for pump fluence of 25, 190 and 330~$\mu$J.cm$^{-2}$.   

For a pump fluence of 25~$\mu$J.cm$^{-2}$, we observe the frequencies that were highlighted in Fig.~\ref{fig:NCCDW}b. However, when the pump fluence increases, we can see a different behavior. In the early time scale, the signal is dominated by the frequencies of the hidden phase: a strong multiple peak component at around 3~THz and a single peak at 4.5~THz (black dashed lines in Fig~\ref{fig:NCCDW}a). Then, after approximately 1~ps, we observe a change of the frequencies towards the ones of the NC phase (white dashed lines in Fig~\ref{fig:NCCDW}a). Moreover, we notice that the frequencies of the early time scales are the same for all power, which confirm that the material undergoes a phase transition to a hidden phase and that the change of frequencies is not due to anharmonicity.

The gated Fourier transforms showns in Fig.~\ref{fig:TFA}c provides similar information. For the early timescale (<1~ps), a broad feature centered at 3~THz and a shoulder at 4.5~THz, which are highlighted by dashed black lines, are seen. However, when the Fourier transforms are performed for time scale greater than 1~ps, the spectra are drastically different, displaying multiple narrow peaks at 2.1 and 3.6~THz, the modes of the NCCDW phase. 

In the time frequency analysis, we also notice the appearance of a dip in the amplitude at the transition between the frequencies of the hidden phase and the NCCDW, highlighted in Fig~\ref{fig:NCCDW}a by white circles. Such dip corresponds to destructive interference, akin to a Fano resonance. A Fano resonance in the frequency domain can be modeled by a phase shift of an oscillatory signal, leading to a discontinuity in the time domain~\cite{PhysRevResearch.3.L032010}. This behavior indicates that a discontinuity occurs when the frequencies change from the H to the NC phase, \textit{i.e.} we have a signature of a phase transition.

Figure~\ref{fig:TFA}b shows the oscillatory part of the transient reflectivity for increasing pump fluence filtered from 3 to 6~THz. The behavior we highlighted in the time frequency analysis can further be observed. At low power, we only observe a single damped oscillation at a frequency of 3.6~THz. When the power increase, the early time scale is modified displaying a faster oscillation, with a frequency of 4.5~THz. The duration of this higher frequency component increases with the deposited energy, and after its disappearance, the system reverts back to oscillations at the frequency of the NC phase.

\section{Discussion}

\begin{figure*}
    \centering
    \includegraphics[width=\textwidth]{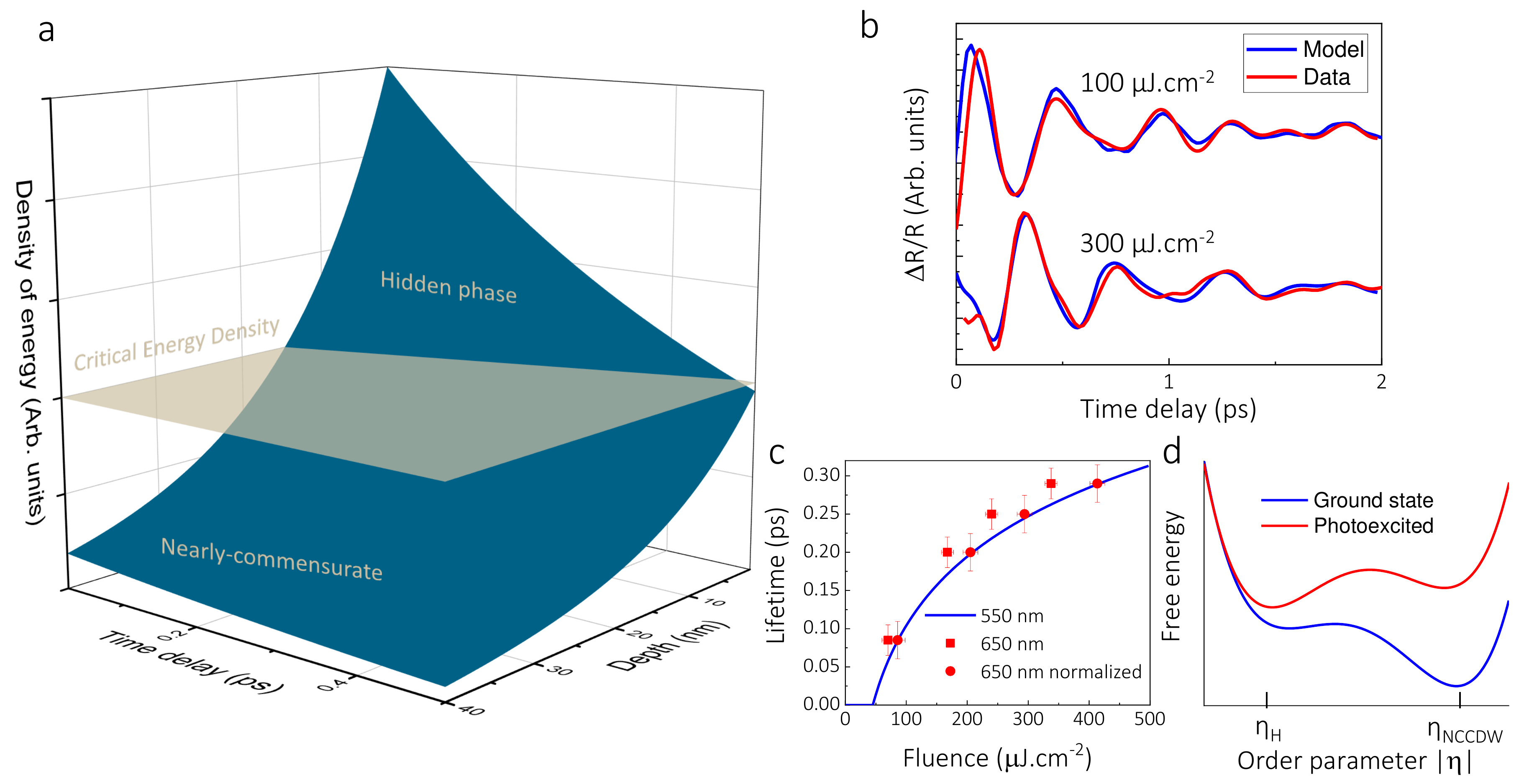}
    \caption{\textbf{Fluence dependence of the phase transition.} (\textbf{a}) Spatiotemporal evolution of the energy deposited by the laser pulse in the system. (\textbf{b}) Oscillatory components of the transient reflectivity for a pump fluence of 100 and 300~$\mu$J.cm$^{-2}$ and fit of these data with the model presented in the text. (\textbf{c}) Lifetime of the hidden phase for different pump fluence based on our model. The lifetime extracted form data obtained with a pump wavelength of 650~nm are also reported, as well as the lifetime at 650-nm, normalized by the number of photoexcited electrons. (\textbf{d}) Schematic representation of the energetic evolution through the phase transition.}
    \label{fig:Lifetime}
\end{figure*}

The observations on the time-frequency analysis allows building a model representative of the phase transition. Here, we provide the main lines of this model, and a detailed derivation is provided in the supplementary note 2. The central idea of our model stems from the observation that the phase transition only occurs for power above 50~$\mu$J.cm$^{-2}$, as seen in Fig.~\ref{fig:NCCDW}b, and therefore, the phase transition requires that the deposited energy density reaches a critical value, $E_c$. We assume a linear absorption of light, which leads to an exponential decay of the deposited energy along the depth of the sample. We also consider that the excess energy deposited by the laser will dissipate on a timescale $\tau$. The overall spatiotemporal profile of the deposited energy thus take the form depicted in Fig.~\ref{fig:Lifetime}a. 

The second step is to model the behaviour of the coherent vibrations during the phase transition. To do so, we consider that while the energy at a given depth is larger than $E_c$, the system oscillates at the frequencies of the hidden phase, and as soon as the energy goes below critical energy, we observe the frequencies of the NC phase. We further need to consider that at the time the phase transition takes place, we need continuity of the atomic position, which provides a boundary condition. This condition is especially important because it explains the Fano-like behavior we observed in the time frequency analysis. Indeed, when the system undergoes the phase transition back to the thermodynamically phase, the equilibrium position around which atoms oscillate is modified. the continuity in the position, leads to a discontinuity in the displacement with regards to the equilibrium position. The transient reflectivity should thus exhibits a discontinuity, which is a necessary condition for the appearance of Fano resonance~\cite{PhysRevResearch.3.L032010}.

The final step of the model is to combine the oscillatory behavior we presented in the previous paragraph, with the energy distribution of Fig.~\ref{fig:Lifetime}a, that shows that the transition does not occur at the same time for different depth within the material. We thus obtain a model with a few fitting parameters, namely, the timescale of energy dissipation, $\tau$, the critical energy density, $E_c$, the relative difference in equilibrium atomic position between the hidden and the NC phase, $\eta_H-\eta_{NCCDW}$, and the amplitude of the various oscillations. We fitted this model to a total of 12 different pump fluences by using a unique set of frequencies and $\tau$, $E_c$, $\eta_H-\eta_{NCCDW}$, while varying the amplitude of the oscillations. Figure~\ref{fig:Lifetime}b shows the oscillatory part of the transient reflectivity obtained for a pump fluence of 100 and 300~$\mu$J.cm$^{-2}$ and fits using the model we developed that shows an excellent agreement. 

As we previously mentioned our fitting procedure leads to a unique set of characteristics of the phase transition. For the energy dissipation timescale, we obtain $\tau$=0.3~ps, which is similar to the electron-phonon scattering time obtained through various experimental methods~\cite{10.1038/nature09539,10.1103/physrevlett.97.067402}. For the critical fluence required to induce the phase transition, we obtain $E_c$=50~$\mu$J.cm$^{-2}$. This value is interesting since it corresponds to a density of photoecited carriers of 5.10$^{-2}$~nm$^{-3}$, similar to the value reported by Han \textit{et al.} to achieve the phase transition from the NCCDW phase to the phase they call the NC$^*$ at room temperature (approximately 0.1~nm$^{-3}$)~\cite{10.1126/sciadv.1400173}. 

These two parameters allows to calculate the lifetime of the hidden phase given by $T=\tau \ln\left(\cfrac{(1-R)F}{E_c \xi}\right)$. Figure~\ref{fig:Lifetime}c shows the lifetime, which corresponds to the time required for the total disappearance of the hidden phase calculated from the parameters extracted by our model. We observe that the lifetime of the hidden phase can be controlled by varying the pump fluence. Such characteristics could find applications in the development of novel photonic devices~\cite{10.1038/nphoton.2015.247}. 

Experiments performed with a pump wavelength of 650~nm leads to different lifetime for similar deposited energy, as shown in Fig.~\ref{fig:Lifetime}c. The lower photon energy at 650~nm means that for a given deposited energy, the number of photoexcited electrons is much larger than for 550~nm. We normalized the lifetime by the density of photoexcited electrons and obtain similar values than for 550~nm. From this observation, we can conclude that the phase transition is triggered by the excitation of a critical density, $n_c$, of electrons with higher kinetic energy. IThe physics of the star of David polarons is ruled by competition between Coulomb repulsion and kinetic energy, \textit{i.e.} the Hubbard model (Eq~1). The excitation of electrons with high kinetic energy leads to an increase of the hopping integral, $t$, and a decrease of the on-site interaction strength, $U$. To accommodate the changes in these interactions, the lattice assumes a different distortion, and a new phase emerges. 

Our model gives access to one last parameter: the difference between the order parameter in the hidden and the NC phase. For CDW, the order parameter corresponds to the displacement of the lattice relative to the undistorted phase. However, we cannot obtain a quantitative value for this parameter since we do not know the proportionality relation between atomic displacement and change in reflectivity. Nevertheless, the fit of our model to experimental data indicates that $|\eta_H|<|\eta_{NCCDW}|$. From this result we can propose a model for the phase transition depicted in Fig.~\ref{fig:Lifetime}d. The system can be described by a double well potential. In the ground state, the system is deformed into SD with the order parameter of the NC phase. In the hidden phase, the energy of NC phase becomes larger than the hidden phase, which leads to the phase transition with the order parameter $\eta_H$. Finally, when the energy relaxes, the energy potential gradually returns to the ground state, and at a time $T$, the NC phase once again becomes more energetically favorable and the system transitions back.

In conclusion, we have investigated the temporal response of the nearly-commensurate charge density waves phase of 1T-TaS$_2$ following the absorption of femtosecond pulses of different fluences. Our analysis reveals that the material undergoes a photo-induced phase transition to a new phase. We show that the phase transition is induced by a change in screening of the on-site Coulomb repulsion due to the excitation of electrons with high kinetic energy. After the electrons lose their kinetic energy through scattering with phonons, the system reverts back to the NC phase. The lifetime of the hidden phase can be controlled by the deposited energy, which will enable the development of new photonic devices. 

\section{Methods}
\subsection{Ultrafast pump-probe spectroscopy}
Experiments were performed using a regeneratively amplified, mode-locked Yb:KGW (Ytterbium-doped potassium
gadolinium tungstate) based femtosecond laser system (Pharos, Light conversion) operating at 1030~nm and delivering pulses of 200~fs at 1~kHz repetition rate. This laser is then used to pump two non-collinearly phase-matched optical parametric amplifiers (NOPAs). A first one (Orpheus-N, Light Conversion), was used to generate pump pulses centered at 550 or 650~nm with pulse duration of about 35 fs. The second NOPA (Orpheus-N, Light Conversion), generated probe pulses at 700~nm with 40~fs pulse duration, that were time delayed with respect to the pump. The pump beam was chopped at the frequency of 500~MHz using a mechanical chopper. Both beams were focused on the sample and the modifications of the
probe reflectivity induced by the pump was time-resolved.

\subsection{1T-TaS$_2$ Sample}
1T-TaS$_2$ samples were purchased from 2d semiconductors (https://www.2dsemiconductors.com/). Samples are grown using the flux zone method which ensure charge density waves behavior. For the data acquisition, the samples were cleaved before experiments.

\section{Data availability}
The authors declare that the data supporting the findings of this study are available from the corresponding authors upon request.

\section{References}
\bibliography{Metastability.bib}

\section{Acknowledgments}

This work was supported by NanoLund, Lund Laser Centre, Crafoordska Stiftelsen, Grant No. 20200630 and Vetenskapsradet, Grant No. 2017-05150

\section{Author contributions}
\end{document}

\title{Supplementary materials for Photo-induced Metastable Phase of 1T-TaS2 with Tunable Lifetime }% Force line breaks with \\

\author{Pierre-Adrien Mante}
\email{pierre-adrien.mante@chemphys.lu.se}
\affiliation{Division of Chemical Physics and NanoLund, Lund University, Sweden}

\author{Chin Shen Ong}
\affiliation{Department of Physics and Astronomy, Uppsala University, Sweden}

\author{Daniel Finkelstein Shapiro}
\affiliation{Instituto de Química, Universidad Nacional Autónoma de México, Mexico}

\author{Arkady Yartsev}
\affiliation{Division of Chemical Physics and NanoLund, Lund University, Sweden}

\author{Oscar Gr\aa n\"as}
\affiliation{Department of Physics and Astronomy, Uppsala University, Sweden}

\author{Olle Eriksson}
\affiliation{Department of Physics and Astronomy, Uppsala University, Sweden}
\affiliation{School of Science and Technology, Örebro University, Sweden}

\date{\today}% It is always \today, today,

\maketitle

\section{Supplementary Figure 1}

\begin{figure}[!h]
    \centering
    \includegraphics[width=\textwidth]{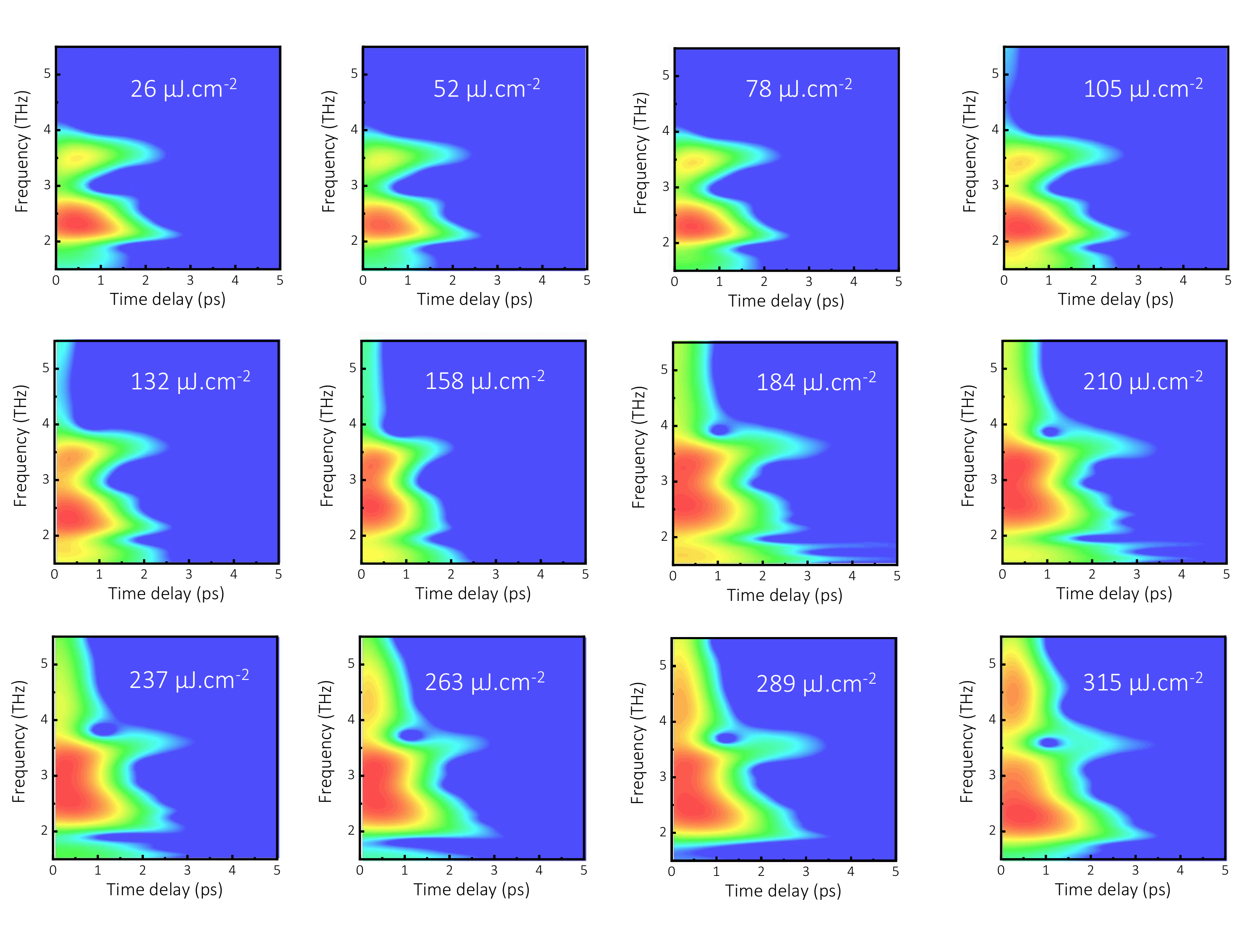}
    \caption{Time frequency analysis of the photo-induced response of 1T-TaS$_{2}$ as a function of pump fluence}
\end{figure}

\section{Supplementary Note 1: Lattice temperature increase}

To estimate the maximum increase of the lattice temperature, we consider the amount of energy deposited in the sample:

\begin{equation}
	\Delta E=\frac{(1-R)F}{\xi}
\end{equation}

Then, considering a lattice specific heat, $C_p$=1.85~J.cm$^{-3}$.K$^{-1}$, we obatin the temperature increase:

\begin{equation}
	\Delta T=\frac{\Delta E}{C_p}
\end{equation}

For the range of fluence considered, we then obtain the temperature increase show in Supplementary Figure~1:

\begin{figure}[!h]
    \centering
    \includegraphics[width=\textwidth]{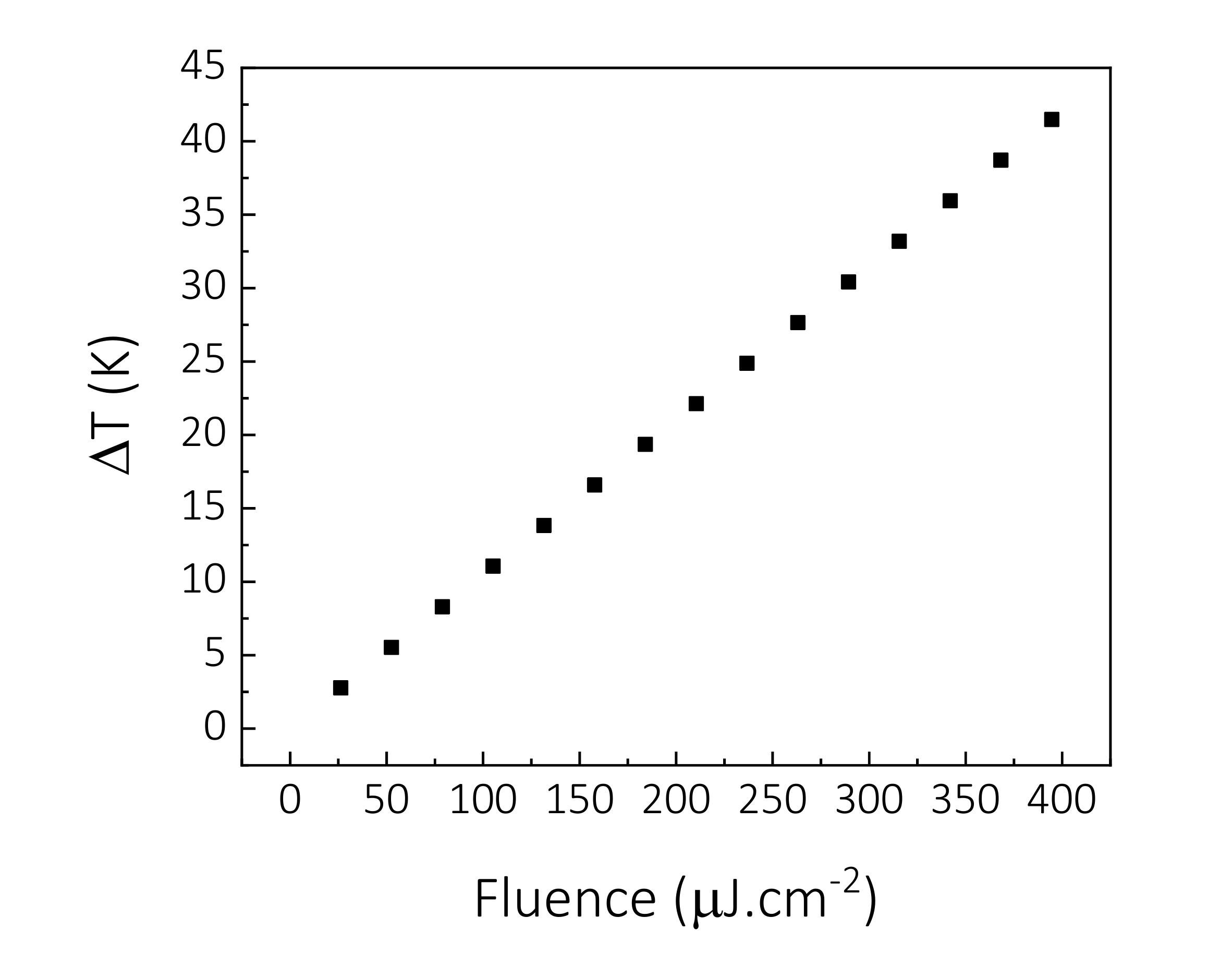}
    \caption{Increase of the lattice temperature for the pump laser fluence investigated}
\end{figure}

The maximum temperature increase is about 41~K, which leads to a lattice temperature of 334~K below the temperature of the NC to IC transition at 350~K.

\section{Supplementary Note 2: Model derivation}

The absorption of a femtosecond pulse with fluence $F$ leads to an increase of the density of energy in the material, $E$, given by: 

\begin{equation}
	E=\frac{(1-R)F}{\xi} e^{-z/\xi}
\end{equation}

where $R$ is the reflectivity coefficient and $\xi$ is the penetration depth at the pump wavelength. Due to thermalization, this excess energy decrease with time, and we obtain the spatiotemporal evolution of the energy density in the material:

\begin{equation}
	E(z,t)=\frac{(1-R)F}{\xi} e^{-z/\xi} e^{-t/\tau}
\end{equation}

where $\tau$ is the typical timescale on which the excess energy dissipate. 

Following our time frequency analysis, we observe that the phase transition can only be observed when the density of energy reaches a critical value $E_c$, which is reached in the material until a critical depth, $z_c=\xi ln(\frac{(1-R)F}{E_c \xi})$ . From this observation, we can calculate for a given fluence, the relative contribution to the signal of the photoexcited materials that undergoes a phase transition:

\begin{align}
W(t)=\frac{e^{-t/\tau}\int_{0}^{z_c}E(z)dz}{\int_{0}^{+\infty}E(z)dz}=&\left\{
\begin{array}{ll}
1-\frac{E_c \xi}{(1-R)Fe^{-t/\tau}}  & \mbox{, if } \frac{(1-R)Fe^{-t/\tau}}{\xi} > E_c \\
0 & \mbox{, if } \frac{(1-R)F}{\xi} < E_c.
\end{array}
\right. 
\end{align}

To understand how the photoinduced transition return back to the NCCDW phase, we start by considering a sample that is homogeneously excited into the new phase, \textit{i.e.} an energy $E>E_c$ is deposited into the material. As we have seen in Eq.~2, this excess energy dissipate on time scale $\tau$, and after a time $t_1=\tau ln(\frac{(1-R)F}{E_c\xi})$, it reverts to the NCCDW phase. This means that for time $t<t_1$, the coherent response will be observed at the frequencies of the hidden phase, for example 4.5~THz, and for $t>t_1$, the observed oscillations will be at the frequencies of the NCCDW phase, which for this mode is at 3.6~THz. We can thus write the evolution of the coherent signal as follows:

\begin{equation}
	S(t)=(1-H(t-t_1))A_1cos(\omega_1t)+H(t-t1)A_3cos(\omega_3(t-t_1))
\end{equation} 

where, $H$ is the Heaviside function. According to previous experimental evidence,~\cite{PhysRevLett.107.177402} we consider that the coherent vibrations are generated through displacive excitation of coherent phonons (DECP), which leads to a cosine response in the hidden phase. At the transition time, $t_1$, the equilibrium atomic position is changed, and the atoms will start oscillating around this novel position in a cosine way. The continuity of atomic position, $X$, gives us one more condition on this equation:

\begin{equation}
	X(t_1)=X_H+A_1cos(\omega_1t_1)=X_{NCCDW}+A_3
\end{equation} 

where $X_H$ and $X_{NCCDW}$ are the equilibrium atomic position in the hidden phase and the NCCDW phase respectively. With this condition, we can rewrite Eq.~4 as follows: 

\begin{equation}
	S(t)=(1-H(t-t_1))A_1cos(\omega_1t)+H(t-t1)[X_{NCCDW}-X_{H}+A_1cos(\omega_1t_1)]cos(\omega_3(t-t_1)
\end{equation} 

The last step in our model is to take into consideration that not all depth of the material are excited with the same initial energy due to absorption of the pump pulse. The overall response is thus the integration of the signal given by Eq.~6 for different transition time dictated by the initial energy deposited in the material. We thus obtain the complete signal given by:

\begin{equation}
	\Delta R(t)=\int_{0}^{t_1}W(t')S(t-(t_1-t'))dt'
\end{equation}

We then use this model to fit the experimental data with a unique set of fitting parameters that are the dissipation time scale, $\tau$, and the critical density of energy, $E_c$, the amplitude of the oscillations and the relative difference in equilibrium position, $X_{NCCDW}-X_{H}$.

\bibliography{Metastability.bib}